\documentclass[conference,onecolumn,10pt]{IEEEtran}
\newtheorem{definition}{Definition}
\usepackage{graphicx}
\usepackage{amsmath}
\usepackage{amssymb}
\usepackage{mathtools}
\usepackage{placeins}
\usepackage[caption=false]{subfig}
\usepackage[hyphens]{url}

\ifCLASSINFOpdf
\else
\fi
\begin{document}
%
\title{Trajectory Generation and Control for Quadrotors}

\author{\IEEEauthorblockN{Sandesh Thapa}
\IEEEauthorblockA{School of Mechanical and Aerospace Engineering\\
Control, Robotics \&  Automation Lab (CoRAL), Fall 2017 \\
Oklahoma State University\\
Stillwater, Oklahoma\\
Email: sandesh.thapa@okstate.edu}}


%


\maketitle

\begin{abstract}

In this project, I explored controller design and trajectory generation for a quadrotor. Differential flatness was used for trajectory generation based on one of the references. A minimum snap trajectory is generated from the given algorithm based on waypoint navigation, and a simple proportional plus derivative (PD) controller is used to track the trajectories. Simulation results are presented which generate smooth trajectories for position and velocity of the quadrotor.

\end{abstract}


%
\IEEEpeerreviewmaketitle

\section{Introduction}
Recently, quadrotors have been widely used for research. Because of their agility in 3D Euclidean space, they can be used for a variety of purposes, including photography, inspection, transportation of loads, geo-surveying, and so on. For a quadrotor maneuvering in a tightly constrained environment, it is necessary to develop optimal trajectories to ensure safe passage through corridors and constraints \cite{mellinger2011minimum}. So, it is necessary to develop flight plans that leverage the dynamics of the system \cite{mellinger2011minimum}.

\section{Project Objective}
The objective of this project is to review and develop an understanding of trajectory generation for quadrotors. I reviewed \cite{mellinger2011minimum} and related sources \cite{mellinger2012trajectory, michael2010grasp} to understand and implement differential flatness for trajectory generation. Example code was given at \cite{AerialRobotics, AerialRobotics1}. I implemented Mellinger's minimum snap trajectory and control algorithm for quadrotors based on differential flatness \cite{mellinger2011minimum, mellinger2012trajectory}. The goal of this project was to develop a strong understanding of quadrotor trajectory generation and control algorithms so that they can be applied to multi-agent path planning and control problems in future work.
\section{Methodology}
In this paper, I used the dynamic model and control from \cite{mellinger2011minimum} and implemented it in MATLAB. I used a PD controller to track a given trajectory. Trajectories are generated based on differential flatness, which ensures smoothness. I compared the minimum snap trajectory and the shortest distance trajectory and distinguished their performance.

\section{Dynamic Model}
\begin{figure}[h!]
\centering
\includegraphics[scale=0.4]{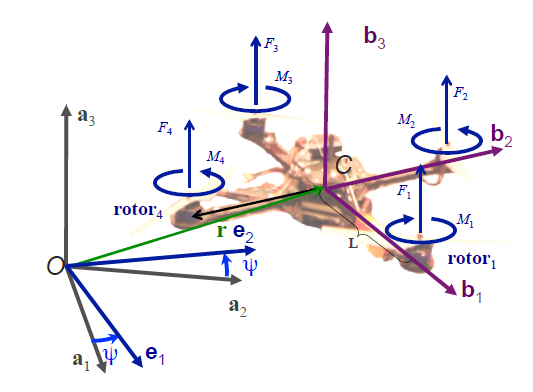}
\caption{Quadrotor model with the body-fixed and inertial reference frames.}
\label{Quadmodel}
\end{figure}
\subsection{Coordinate Systems}
The coordinate systems with a free body diagram for the quadrotor are shown in Fig.~\ref{Quadmodel}. $\mathcal{A}$ is the inertial frame with axes $\mathsf{a}_1$, $\mathsf{a}_2$ and $\mathsf{a}_3$. The body frame $\mathcal{B}$ is attached to the center of mass of the quadrotor, which is coincident with the inertial frame. $\mathcal{C}$ is the center of mass of the rotor and ${L}$ is the length of the rotors from the center $\mathcal{C}$.

\subsection{Motor Model}
Each rotor has an angular speed $\omega_i$ and produces a vertical force $F_i$, which is given by
\begin{equation}
\label{Fi}
F_i = k_F\omega_i^2,
\end{equation}
$k_F \approx 6.11\times10^{-8} \frac{N}{rmp^2}$ \cite{michael2010grasp}.
\begin{equation}
\label{Mi}
M_i = k_M\omega_i^2,
\end{equation}
$k_M \approx 1.5\times10^{-9} \frac{Nm}{rmp^2}$ \cite{michael2010grasp}.
\subsection{Rigid Body Dynamics}
The orientation of the quadrotor is described by a triplet of yaw-pitch-roll ($\phi$, $\theta$ and $\psi$) $Z-Y-X$ Euler angles. The position of the center of mass of the quadrotor is given by
\begin{equation}
\textbf{r} = \begin{bmatrix}
x&y&z
\end{bmatrix}^T.
\end{equation}
The state of the quadrotor is given by
\begin{equation}
q = \begin{bmatrix}
x & y & z & \phi & \theta & \psi
\end{bmatrix}^T.
\end{equation}
The rotation matrix from $\mathcal{B}$ to $\mathcal{A}$ is given by,
\begin{equation}
\label{Rot}
R_{BA} = \begin{bmatrix}
c\phi c\theta-s\phi s\psi s\theta & -c\phi s\psi & c\psi s\theta + c \theta s \phi s \psi \\
c \theta s \psi + c \psi s \phi s \theta & c \phi c \psi & s \psi s \theta -c \psi c \theta s \phi \\
-c \phi s \theta & s \phi & c \phi c \theta
\end{bmatrix}.
 \end{equation}
 The angular velocity of the robot in the body frame is denoted as $q$, $q$ and $r$. These values are related to derivatives of the roll, pitch and yaw angles according to
 \begin{equation}
 \label{pqr}
 \begin{bmatrix}
 p\\
 q\\
 r\\
 \end{bmatrix} = \begin{bmatrix}
 c\theta & 0 & -c\phi s \theta \\
 0 & 1 & s \phi \\
 s \theta & 0 & c \phi c \theta
 \end{bmatrix} \begin{bmatrix}
 \dot{\phi}\\
 \dot{\theta}\\
 \dot{\psi}
 \end{bmatrix}.
 \end{equation}

\subsubsection{Newton's Equations of Motion}
The forces in the system are gravity and the forces from each rotor $F_i$. The equation governing the acceleration of the center of mass of the quadrotor is
\begin{equation}
\label{nweq}
m\ddot{\textbf{r}} = \begin{bmatrix}
0 \\
0\\
mg \\
\end{bmatrix} + R_{BA} \begin{bmatrix}
0 \\
0\\
F_1 + F_2 + F_3 + F_4
\end{bmatrix},
\end{equation}
The first input to the system $u_1$ is defined as
$$u_1 = \sum_{i=1}^4F_i$$.
\subsubsection{Euler's Equations of Motion}
In addition to forces, each rotor produces a moment perpendicular to the plane of rotation of the blade, $M_i$. Rotors 1 and 3 rotate in the $-\mathsf{b}_3$ direction while 2 and 4 rotate in the $+\mathsf{b}_3$ direction. The moment produced on the quadrotor is opposite to the direction of rotation of the blades. So, $M_1$ and $M_3$ act in the $\mathsf{b}_3$ direction while $M_2$ and $M_4$ act in the $-\mathsf{b}_3$ direction. The angular acceleration is determined by,
\begin{equation}
\label{ELEq}
I\begin{bmatrix}
\dot{p}\\
\dot{q} \\
\dot{r}
\end{bmatrix} = \begin{bmatrix}
L(F_2-F_4) \\
L(F_3-F_1)\\
M_1-M_2+M_3-M_4
\end{bmatrix} - \begin{bmatrix}
p\\
q\\
r\\
\end{bmatrix}\times I  \begin{bmatrix}
p\\
q\\
r\\
\end{bmatrix}.
\end{equation}
We can rewrite this as:
\begin{equation}
\label{ELEq}
I\begin{bmatrix}
\dot{p}\\
\dot{q} \\
\dot{r}
\end{bmatrix} = \begin{bmatrix}
0 & L & 0 & -L \\
-L & 0 & L & 0 \\
\gamma & -\gamma & \gamma & -\gamma
\end{bmatrix}\begin{bmatrix}
F_1\\
F_2\\
F_3\\
F_4 \\
\end{bmatrix} - \begin{bmatrix}
p\\
q\\
r\\
\end{bmatrix}\times I  \begin{bmatrix}
p\\
q\\
r\\
\end{bmatrix},
\end{equation}
where $\gamma = \frac{k_M}{k_F}$. The second input $\textbf{u}_2$ is defined as: $$ \textbf{u}_2 =
\begin{bmatrix}
0 & L & 0 & -L \\
-L & 0 & L & 0 \\
\gamma & -\gamma & \gamma & -\gamma
\end{bmatrix}  \begin{bmatrix}
F_1\\
F_2\\
F_3\\
F_4 \\
\end{bmatrix}.$$

\section{Control}
The controller used in \cite{mellinger2011minimum, michael2010grasp} is derived by linearizing the equation of motion about an operating point, $\textbf{r} = \textbf{r}_0$, $\theta = \phi = 0$, $\psi = \psi_0$, $\dot{\textbf{r}} = 0$, and $\dot{\phi}= \dot{\theta} = \dot{\psi} = 0$. Using a small angle approximation for roll and pitch angles ($c\phi = 1$, $c\theta = 1$, $s\phi \approx \phi$, and $s\theta \approx \theta$) gives us the following linearized equations:
\begin{equation}
F_{i,0} = \frac{mg}{4},
\end{equation}
At the nominal state, the inputs for hover are $u_{1,0} = mg$, $\textbf{u}_{2,0} = 0$.

Linearizing \eqref{nweq}, we get:
\begin{equation}
\begin{aligned}
\label{rddot}
\ddot{r}_1 &= g(\Delta\theta \cos \psi_0 + \Delta \phi \sin \psi_0) \\
\ddot{r}_2 &= g(\Delta\theta \sin \psi_0 - \Delta \phi \cos \psi_0) \\
\ddot{r}_3 &= \frac{1}{m}u_1 - g
\end{aligned}
\end{equation}
Linearizing \eqref{ELEq}, we get:
\begin{equation}
\label{pqrddot}
\begin{bmatrix}
\dot{p} \\
\dot{q}\\
\dot{r} \\
\end{bmatrix} = I^{-1}
\begin{bmatrix}
0 & L & 0 & -L \\
-L & 0 & L & 0 \\
\gamma & -\gamma & \gamma & -\gamma
\end{bmatrix}  \begin{bmatrix}
F_1\\
F_2\\
F_3\\
F_4 \\
\end{bmatrix}
\end{equation}
The rotor craft is assumed to be symmetric so $I_{xx} = I_{yy}$,
\begin{equation}
\begin{aligned}
\dot{p} &= \frac{u_{2,x}}{I_{xx}} = \frac{L}{I_{xx}}(F_2-F_4)\\
\dot{q} &= \frac{u_{2,y}}{I_{yy}} = \frac{L}{I_{yy}}(F_3-F_1)\\
\dot{r} &= \frac{u_{2,z}}{I_{zz}} = \frac{\gamma}{I_{zz}}(F_1-F_2+F_3-F_4)
\end{aligned}
\end{equation}
\subsection{Position and Attitude Controller}
\begin{figure}[h!]
\centering
\includegraphics[scale=0.5]{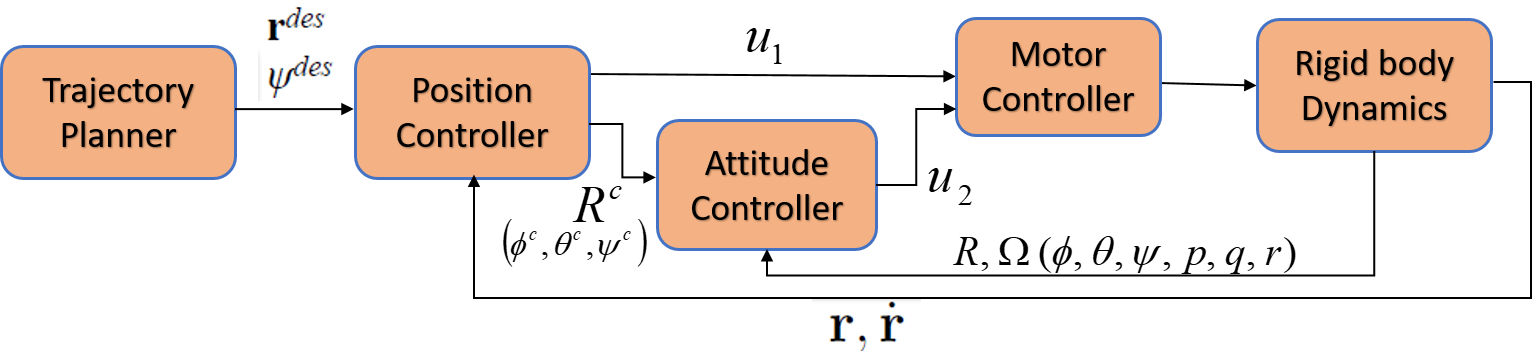}
\caption{The position and attitude control loops.}
\label{Control1}
\end{figure}

The control problem is to determine the four inputs, ($u_1, \textbf{u}_2$) required to hover or to follow a desired trajectory, $\textbf{z}_des$. As shown in Fig. \ref{Control1}, the error in the robot's position is used to derive the controller from \eqref{rddot}. The equations in \eqref{rddot} also allow us to derive a desired orientation. The attitude controller for this desired orientation is derived from \eqref{pqrddot}. We require the attitude control loop to run at a higher rate than the position control loop.
\subsubsection{Attitude Control}
Mellinger \cite{mellinger2011minimum} developed the following PD controller for attitude control:
\begin{equation}
\label{attitudeC}
\textbf{u}_2 = \begin{bmatrix}
k_{p,\phi}(\phi_{des} - \phi)+ k_{d,\phi}(p_{des}-p) \\
k_{p,\theta}(\theta_{des} - \theta)+ k_{d,\theta}(q_{des}-q) \\
k_{p,\psi}(\psi_{des} - \psi)+ k_{d,\psi}(r_{des}-r)
\end{bmatrix}.
\end{equation}
Here, $k_{p,i} \quad \text{and} \quad k_{d,i}$ are the control gains that need to be tuned.
\subsubsection{Position Control}
If a desired trajectory is given as: $$\textbf{z}_des = \begin{bmatrix}
\textbf{r}_{des}(t) \\
\psi_T(t)
\end{bmatrix},$$
the following PD control is used to track the yaw angle and the three position coordinates of the rotor.
\begin{equation}
\ddot{r}_{i,c} = \ddot{r}_{i,des} + k_{d,i}(\dot{r}_{i,des} - \dot{r}_i ) + k_{p,i}(r_{i,des}-r_i).
\end{equation}
which gives us
\begin{equation}
\label{desddt}
u_1 = m(g+\ddot{r}_{3,c}).
\end{equation}

We find the desired roll and pitch angles as
\begin{equation}
\begin{aligned}
\phi_{des} &= \frac{1}{g}(\ddot{r}_{1,c}\sin \psi_{des} - \ddot{r}_{2,c}\cos \psi_{des} ), \\
\theta_{des} &= \frac{1}{g}(\ddot{r}_{1,c}\cos \psi_{des} + \ddot{r}_{2,c}\sin \psi_{des}. )
\end{aligned}
\end{equation}
The desired roll and pitch velocities are taken to be zero.
\section{Differential Flatness}
\begin{definition}
A nonlinear system
\begin{equation}
\label{xdot}
\begin{aligned}
\dot{x} &= f(x,u), & x \in \mathbb{R}^n, & \quad u \in \mathbb{R}^m,\\
y &= h(x),   & y \in \mathbb{R}^m, &
\end{aligned}
\end{equation}
\end{definition}
is differentially flat if we can find outputs $z \in \mathbb{R}^m$ of the form
\begin{equation}
\label{z}
z \ \zeta(x,u,\dot{u},...,u^{(l)}),
\end{equation}
such that
\begin{equation}
\begin{aligned}
x &= x(z ,\dot{z}, ...,z^{(l)} \coloneqq x(\bar{z}),\\
u &= u(z,\dot{z},...,z^{(l)}) \coloneqq u(\bar{z}).
\end{aligned}
\end{equation}
A nonlinear system is differentially flat if we can find a set of outputs such that we can express all states and inputs in terms of those outputs and their derivatives \cite{van1997real}.
\subsection{Example}
\begin{figure}[h!]
\centering
\includegraphics[scale=0.4]{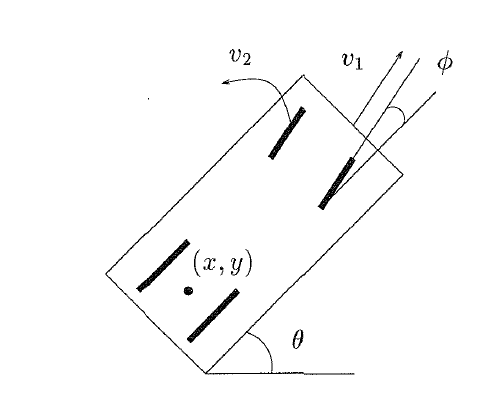}
\caption{Kinematic Car}
\label{Car}
\end{figure}
A kinematic car is a differentially flat system, as shown in \cite{van1997trajectory} \\
Equations of motion
\begin{equation}
\begin{aligned}
\dot{x} &= \cos \theta \cos \phi v_1 \\
\dot{y} &= \sin \theta \cos \phi v_2 \\
\dot{\theta} &= \frac{1}{l}\sin \phi v_1 \\
\phi &= v_2
\end{aligned}
\end{equation}
where $v_1$ and $v_2$ are the inputs of the system.
If the output is chosen to be $(x, y)$, $\phi$ can be expressed in terms of the flat outputs as
\begin{equation}
\phi \arctan (\frac{1}{l}({\dot{x}^2+\dot{y}^2)}^{\frac{3}{2}},\dot{x}\ddot{x}-\dot{y}\ddot{y}),
\end{equation}
which shows that the kinematic car is a differentially flat system.
\subsection{Quadrotor as a Differentially Flat System}
From \eqref{nweq}, we get
\begin{equation}
\label{nweq2}
m\ddot{\textbf{r}} = \begin{bmatrix}
0 \\
0\\
mg \\
\end{bmatrix} + R_{BA} \begin{bmatrix}
0 \\
0\\
u_1
\end{bmatrix}
\end{equation}
Using the linearized equations of motion near hover with small angle assumptions, we get,
\begin{equation}
\label{Dfu1}
\begin{aligned}
m\ddot{x} &= (\theta c \psi + \phi s \psi)u_1 \\
m\ddot{y} &= (\theta s \psi - \phi c \psi)u_1 \\
m\ddot{z} &= -mg + u_1
\end{aligned}
\end{equation}
It shows that the second derivative of position is proportional to the input $u_1$.

From the angular rate equation \eqref{pqr}, we get:
\begin{equation}
\label{DFu2}
\begin{aligned}
p &= \dot{\phi}c\theta - \dot{\psi}c \phi s \theta  \\
q & = \dot{\theta} + \dot{\psi}s \phi \\
r & = \dot{\phi} s \theta + \dot{\psi}c\phi c \theta
\end{aligned}
\end{equation}
Substituting the approximation:

$\sin(\theta) \approx \theta, \sin(\phi) \approx \phi, \cos(\phi) = \cos(\theta) \approx 1$, we get:
\begin{equation}
\label{DFu2qpqr}
\begin{aligned}
p &= \dot{\phi} - \dot{\psi} \theta  \\
q & = \dot{\theta} + \dot{\psi}\phi \\
r & = \dot{\phi} \theta + \dot{\psi}
\end{aligned}
\end{equation}
Further substituting:

$\dot{\psi} \theta \approx \dot{\psi}\phi \approx \dot{\phi} \theta  = 0  $,
we get:
\begin{equation}
\label{jpayetehi}
\begin{aligned}
p = \dot{\phi} \\
q = \dot{\theta} \\
r = \dot{\psi}
\end{aligned}
\end{equation}
Now, considering only the principal axes of inertia of the quadrotor and using \eqref{ELEq}, we get:
\begin{equation}
\label{bigeqn}
\begin{bmatrix}
I_{xx} & 0 & 0\\
0 & I_{yy} & 0 \\
0 & 0 & I_{zz} \\
\end{bmatrix} \begin{bmatrix}
\dot{q} \\
\dot{q} \\
\dot{r} \\
\end{bmatrix} = \begin{bmatrix}
u_{2,x} \\
u_{2,y} \\
u_{2,z} \\
\end{bmatrix} - \begin{bmatrix}
0 & r & -q\\
-r& 0 & p \\
 q & -p & 0 \\
\end{bmatrix} \begin{bmatrix}
I_{xx} & 0 & 0\\
0 & I_{yy} & 0 \\
0 & 0 & I_{zz} \\
\end{bmatrix} \begin{bmatrix}
p \\
q \\
r \\
\end{bmatrix}
\end{equation}
From \eqref{bigeqn}, we get:
\begin{equation}
\begin{aligned}
I_{xx}\dot{p} & = u_{2,x} - I_{yy}qr + I_{zz} qr \\
I_{yy}\dot{q} & = u_{2,y} + I_{xx}pr - I_{zz} pr \\
I_{xx}\dot{r} & = u_{2,z} - I_{xx}pq + I_{zz} pq \\
\end{aligned}
\end{equation}

Using the approximation, $\dot{\psi} \dot{\theta} \approx \dot{\psi}\dot{\phi} \approx \dot{\phi} \dot{\theta} = pq = pr = qr = 0   $,
\begin{equation}
\label{eqn:mm}
\begin{aligned}
I_{xx}\dot{p} & = u_{2,x}  \\
I_{yy}\dot{q} & = u_{2,y}  \\
I_{xx}\dot{r} & = u_{2,z} \\
\end{aligned}
\end{equation}

From \eqref{jpayetehi} and \eqref{muji}, we get:
\begin{equation}
\label{AngDDot}
\begin{aligned}
\ddot{\phi} & = \frac{u_{2,x}}{I_{xx}} \\
\ddot{\theta} &= \frac{u_{2,y}}{I_{yy}}  \\
\ddot{\psi} & = \frac{u_{2,z}}{ I_{xx}}
\end{aligned}
\end{equation}
Differentiating \eqref{Dfu1},
\begin{equation}
\label{X3Dot}
\dddot{x} = \Big(\theta c \psi + \phi s \psi)\Big)\dot{u_1}+ \Big(\dot{\theta} c \psi -\theta s\psi \dot{\psi} + \dot{\phi} s \psi + \phi c \psi \dot{\psi}\Big)u_1
\end{equation}

Differentiating again,

 \begin{equation}
\label{X4Dots}
\ddddot{x} = \Big(\theta c \psi + \phi s \psi)\Big)\ddot{u_1}+ 2\Big(\dot{\theta} c \psi -\theta s\psi \dot{\psi} + \dot{\phi} s \psi + \phi c \psi \dot{\psi}\Big)\dot{u_1} + \\
\Big(\ddot{\theta}c\psi -\dot{\theta}s \psi \dot{\psi} - \theta s \psi \ddot{\psi}  - \theta s \psi \dot{\psi}^2 + \ddot{\phi}s \psi + \dot{\phi} c \psi \dot{\psi} +  \dot{\phi} c \psi \ddot{\psi} - \dot{\phi} c \psi \dot{\psi}^2 \Big)u_1
\end{equation}
After substituting $\ddddot{x}$ into \eqref{eqn:mm}, we can prove that the fourth derivative of position is proportional to $u_2$. So, it is proved that the quadrotor is a differentially flat system \cite{mellinger2011minimum} \cite{AerialRobotics}.

\section{3D Trajectory Generation}
For the purpose of this simulation, I used the optimal trajectory defined in \cite{mellinger2011minimum}.
For the optimal trajectory, the following function is used based on Calculus of Variations,
\begin{equation}
\label{opt}
x*(t) = arg \min_{x(t)} \int_0^TL(\dot{x},xt)dt
\end{equation}
Solving the functional $L$ using the Euler-Lagrange equation gives us the optimal function.
\begin{equation}
{d}{dt}\Big(\frac{\partial L}{\partial \dot{x}}\Big) - \frac{\partial L}{\partial {x}} = 0
\end{equation}

For smooth trajectory generation, Mellinger \cite{mellinger2011minimum} used the following functional
\begin{equation}
\label{opt}
x*(t) = arg \min_{x(t)} \int_0^T(({x^n})^2)dt
\end{equation}
When n = 1, we get the shortest distance with minimum velocity, n = 2 gives minimum acceleration, n = 3 gives minimum acceleration, and n = 4 produces the minimum snap trajectory. For the rest of the paper, I will use the minimum snap trajectory for simulation purposes.

\subsection{Minimum Snap Trajectory}

A minimum snap trajectory is a 7th order polynomial of time \cite{AerialRobotics}, \cite{mellinger2011minimum}. If we are given a set of $n+1$ waypoints $w_0,...w_n$, the minimum snap trajectory is a piecewise polynomial composed of $n$ 7th order polynomials. Each polynomial piece $p_i$ travels between a pair of waypoints $w_{i-1}$ and $w_i$ and takes a known amount of time $T_i$ to complete $i = 1,...,n$.

Let $S_0 = 0$ and for $i = 1,...,n, S_i = \sum_{k=1}^{i}T_k$. $S_i$ is the time it takes to reach waypoint $w_i$ from waypoint $w_0$. Then the polynomial $p_i$ has the form:
\begin{equation}
p_i(t) = \alpha_{i0} + \alpha_{i1}\frac{t-S_{-i1}}{T_i} +\alpha_{i2}\Big(\frac{t-S_{-i1}}{T_i}\Big)^2 + ...+ \alpha_{i7}\Big(\frac{t-S_{-i1}}{T_i}\Big)^7
\end{equation}
To obtain the complete equation for the piecewise trajectory, we need to solve for all the coefficients $\approx_{ij}$. There are $8n$ such coefficients. These coefficients must satisfy a series of constraints. First, the polynomial must go through all the waypoints
\begin{equation}
p_i(S_{i-1}) = w_{i-1} \quad \text{and} \quad p_i(S_i) = w_i \quad \text{for all i = 1,...,n (2n constraints)}
\end{equation}

Second, the velocity, acceleration, and jerk are zero at the endpoints \\
\begin{equation}
p_1^{(k)}(S_{0}) = p_n^{(k)} = 0 \quad \text{and} \quad \text{for all k = 1,.., 3 (6 constraints)}
\end{equation}
Third, the velocity, acceleration, and 3rd to 6th derivatives are continuous
\begin{equation}
p_1^{(k)}(S_{i}) = p_{i+1}^{(k)} \quad \text{and} \quad \text{for all k = 1,.., 6 (6n-6 constraints)}
\end{equation}
The coefficients can be solved for by converting the equations into matrix form
\begin{equation}
A\alpha = b
\end{equation}
The coefficients can be obtained by solving
\begin{equation}
\alpha = A^{-1}b
\end{equation}

\section{Simulation Results}
In this section, results obtained from minimum snap trajectory generation are included. Example code was given in \cite{AerialRobotics} \cite{AerialRobotics1}, and I implemented the trajectory generation and control algorithm in their example code. The model used is the AscTec Hummingbird quadrotor. The mass of the rotor is 0.18 kg, the length (L) of the rotor from center of mass to motors is 0.086 m, and the principal moments of inertia are $I_{xx} = 0.00025 kgm^2$, $I_{yy} = 0.000232 kgm^2$, $I_{zz} = 0.0003738 kgm^2$.
\begin{figure}[h!]
\centering
\includegraphics[scale=0.2]{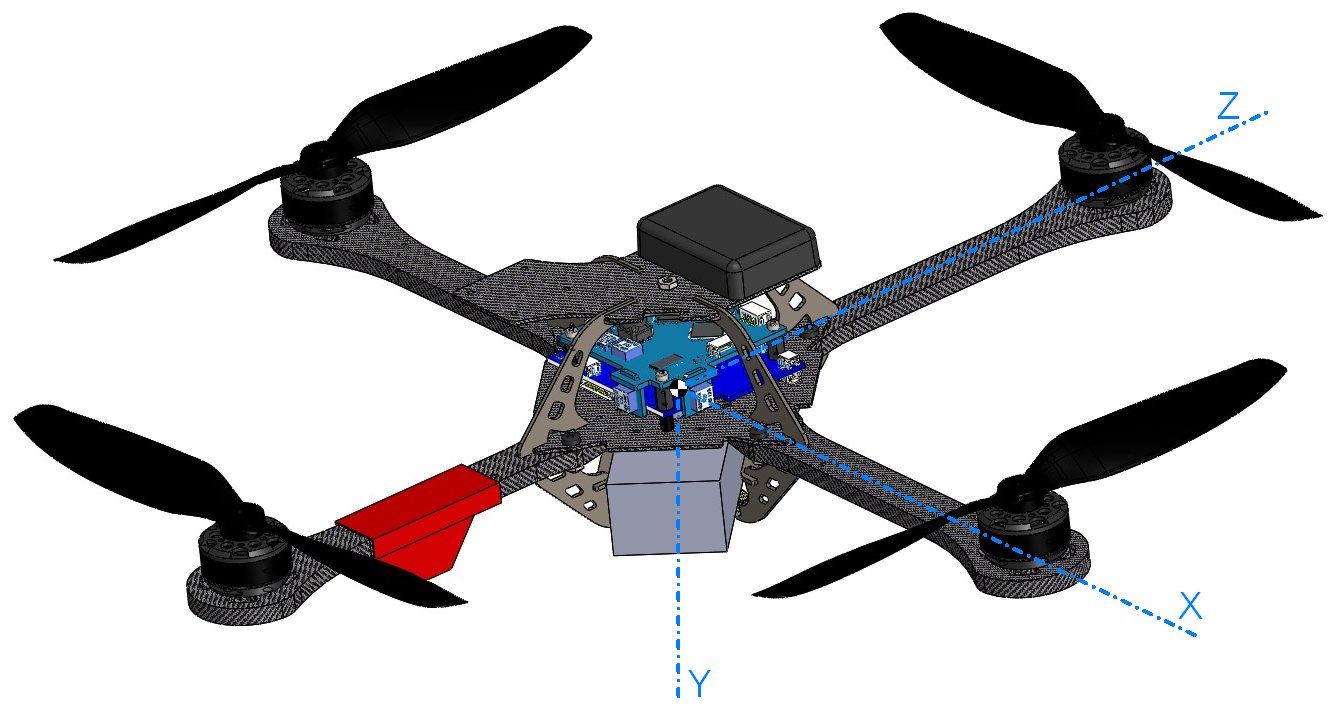}
\caption{AscTec Hummingbird quadrotor used in the paper for simulation purposes \cite{mellinger2011minimum} \cite{wiki1}}
\label{Quadmodel}
\end{figure}
\begin{figure*}[htpb]
\centering
\subfloat[]{\includegraphics[width=3.5in]{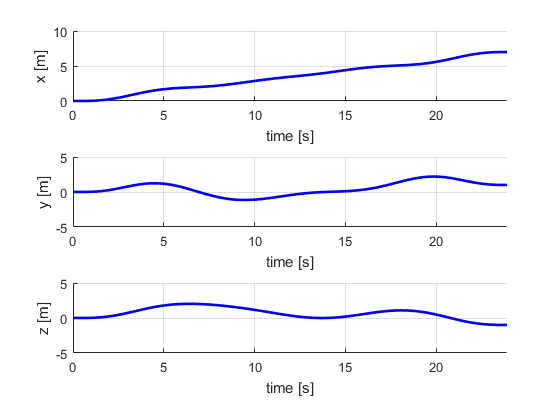}%
\label{fig_first_case}}
\hfil
\subfloat[]{\includegraphics[width=3.5in]{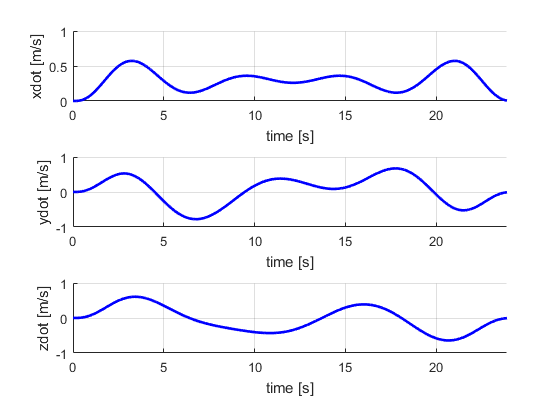}%
\label{fig_second_case}}
\caption{Position and velocity trajectories for the quadrotor based on minimum snap. All the trajectories are smooth and based on differential flatness.}
\label{fig_sim}
\end{figure*}

\begin{figure*}[htpb]
\centering
\subfloat[]{\includegraphics[width=3.5in]{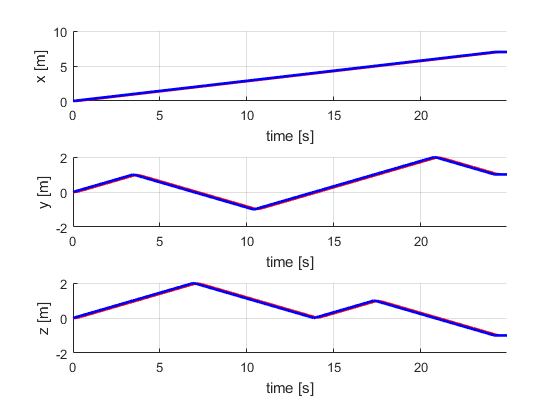}%
\label{fig_first_case1}}
\hfil
\subfloat[]{\includegraphics[width=3.5in]{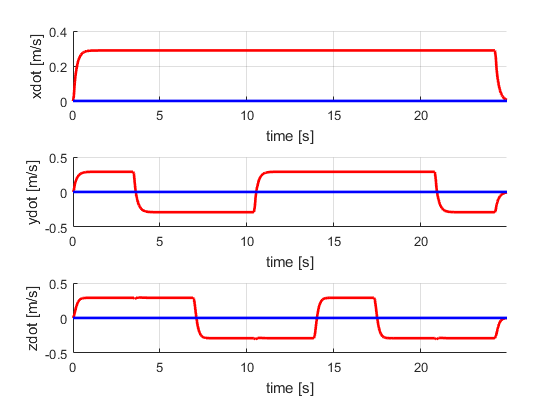}%
\label{fig_second_case1}}
\caption{Position and velocity trajectories for the quadrotor without minimum snap. The points are connected linearly with each other based on the shortest distance (minimum velocity).}
\label{fig_sim1}
\end{figure*}

\FloatBarrier
As we can observe from Fig.~\ref{fig_sim}, the trajectories are smooth and there is no discontinuity in the system. The smooth trajectories allow better performance and planning of the quadrotor. However, in Fig.~\ref{fig_sim1}, it can be observed that the trajectories are not continuous; they pass through the specified waypoints, but because of the sudden jumps in the velocity and position graphs, they are not smooth. Also, because of the inertia present in the system, it may not be feasible to make such a sharp turn.



\section{Conclusion}
From this project, I learned an interesting application of trajectory generation for quadrotors. Minimum snap trajectories can be used for aggressive maneuvers and perching, and the system will still have no perturbations. This is one of the important concepts I learned from this project. The trajectories are always smooth and optimal. One can select waypoints and introduce more dynamic behavior to the quadrotor. I plan to use differential flatness based trajectory generation for my future work and test it on the experimental platform with one of our rotors. I also plan to extend it to multi-agent motion planning and control.

%
%
%
%
%




%

\bibliographystyle{IEEEtran}
\bibliography{IEEEabrv,mybib}


\end{document}